\documentclass[12pt]{iopart}

\usepackage{graphicx}
\usepackage{bm}
\usepackage{iopams}

\begin{document}


\title{ Majority carrier type inversion in FeSe family and ``doped semimetal''  scheme in iron-based superconductors}

\author{Y.A.~Ovchenkov$^{1*}$, D.A.~Chareev$^{2,3.4}$, V.A.~Kulbachinskii$^{1,5}$, \\ V. G.~Kytin$^{1,6}$, S.V.~Mishkov$^1$, D. E.~Presnov$^7$, \\O.S.~Volkova$^{1,3,8}$  and A.N.~Vasiliev$^{1,8,9}$}

\address{$^1$Faculty of Physics, M.V. Lomonosov Moscow State University, Moscow 119991, Russia}

\address{$^2$Institute of Experimental Mineralogy, RAS, Chernogolovka, 123456, Russia}
\address{$^3$Ural Federal University, Ekaterinburg  620002 , Russia}
\address{$^4$Institute of Geology and Petroleum Technologies, Kazan Federal University, Kazan 420008 Russia}
\address{$^5$Moscow Engineering Physics Institute (MEPhI), National Nuclear Research University, Moscow 115409, Russia}
\address{$^5$VNIIFTRI, Mendeleevo, Moscow region 141570, Russia}
\address{$^7$Skobeltsyn Institute of Nuclear Physics, Moscow 119991, Russia}
\address{$^8$National University of Science and Technology 'MISiS' , Moscow 119049, Russia}
\address{$^{9}$National Research South Ural State University, Chelyabinsk 454080 , Russia}

\ead{$^*$ovtchenkov@mig.phys.msu.ru}

\begin{abstract}
The field and temperature dependencies of the longitudinal and Hall resistivity have been studied for high-quality FeSe${}_{1-x}$S${}_{x}$ (x up to 0.14) single crystals. Quasiclassical analysis of the obtained data indicates a strong variation of the electron and hole concentrations under the studied isovalent substitution and proximity of FeSe to the point of the majority carrier-type inversion. On this basis, we propose a ``doped semimetal'' scheme for the superconducting phase diagram of the FeSe family, which can be applied to other iron-based superconductors. In this scheme, the two local maxima of the superconducting temperature can be associated with the Van Hove singularities of a simplified semi-metallic electronic structure. The multicarrier analysis of the experimental data also reveals the presence of a tiny and highly mobile electron band for all the samples studied. Sulfur substitution in the studied range leads to a decrease in the number of mobile electrons by more than ten times, from about 3\% to about 0.2\%. This behavior may indicate a successive change of the Fermi level position relative to singular points of the electronic structure which is consistent with the ``doped semimetal'' scheme. The scattering time for mobile carriers does not depend on impurities, which allows us to consider this group as a possible source of unusual acoustic properties of FeSe.
\end{abstract}




\maketitle


\section{Introduction}
The energy of the superconducting pairing increases with the number of interacting particles. Therefore, possible enhancement of superconductivity by increasing the electron density in the case where the Fermi level is near Van Hove singularities has been studied starting with V${}_{3}$X type compounds \cite{PhysRevLett.19.1039}. The modification of the pairing at these singular points is also studied \cite{PhysRevLett.56.2732} for a long time.

The discovery of cuprate superconductors motivated an in-depth study of the Van Hove scenario for high-temperature superconductivity \cite{EPL-3-1225,JETP.46.118,PhysRevB.45.5714} (HTSC). Because of continued support by new experimental facts, the Van Hove scenario remains relevant to cuprate HTSC for many years \cite{ABRIKOSOV200097, PhysRevLett.89.076401}.

For iron-based superconductors (IBS), many phenomena have also been discovered which are in good agreement with the Van Hove scenario of HTSC. Moreover, it has been suggested that density waves or their fluctuations which are considered responsible for superconductivity in many families of IBS are a consequence of tuning to the nearest Van Hove instability in multiband materials \cite{PhysRevLett.105.067002}.

Among IBS, iron selenide possesses superconductivity in almost stoichiometric form \cite{2017_Coldea, Bohmer2018}, which makes it possible to study the electronic properties of IBS on structurally perfect crystals. For this material, several interesting phenomena have been discovered that can be related to the properties of electrons near singular points of the electronic structure or in small pockets of the Fermi surface. The most interesting phenomenon is the absorption of the C${}_{11}$  acoustic mode at a record level for metals \cite{epl_101-5-56005}. This can be regarded as the experimental detection of a giant electron-phonon interaction for some small electron or hole pockets. It is also interesting that the angle-resolved photoemission study of the substituted FeSe showed a significant change in the superconducting gap between Fermi surface pockets \cite{PhysRevLett.117.157003}. All this suggests that there may be a small group of carriers with ``enhanced'' superconducting properties.

A group of carriers with distinguishing properties have long been noticed in the transport properties of some IBS \cite{PhysRevB.80.140508}. A quasiclassical analysis of transport properties indicates the presence of a small group of mobile carriers in many IBS, including FeSe. The origin of these carriers has not yet been established. The appearance of this group may be a consequence of the Fermi velocity and the effective mass modification for a small Fermi pocket or, alternatively, due to the formation of the Dirac cone \cite{PhysRevB.90.144516}. We suppose that mobile carriers can originate from a small anisotropic pocket near the Van Hove point which is formed in the orthorhombic phase\cite{SUST-30-3-035017}.

Here we present the experimental magnetotransport data for high-quality FeSe${}_{1-x}$S${}_{x}$ crystals and results of quasiclassical multicarrier analysis of these data. The quasiclassical analysis reveals a minority highly mobile band in all studied crystals. The carrier concentration in this band does not exceed a few percents and rapidly decreases with increasing substitution or impurity level. The scattering time for this carriers does not depend on the impurity level that perhaps explains the detected record values in acoustic properties.

The properties of the main carriers also strongly change under substitution, suggesting the possibility of the majority carrier-type inversion in FeSe family. The carrier-type inversion makes it possible to consider a model of ``doped semimetal'' for the phase diagram of the iron-based superconductors in an analogy of the model of ``doped Mott insulator''  for superconducting phase diagrams of cuprates. Thus, both phase diagrams of HTSC can be described consistently under the assumption that the principal factor is the position of the Fermi level relative to the bands' extrema.

\section{Experiment}
The crystals of FeSe${}_{1-x}$S${}_{x}$  were prepared using conventional KCl/AlCl${}_{3}$ flux technique  \cite{Chareev2016, CrystEngComm20.2449}. The energy dispersive microanalysis confirms a good chemical homogeneity of the grown crystals. The chemical composition was studied at three points for four average size crystals from each growth batch, which provided the statistical error in sulfur content lower than 5\% for all batches. 

Magnetoresistance and Hall effect measurements were done using the EDC option of MPMS 7T with Keithley 2400 and Keithley 2192. Electrical measurements were done on cleaved rectangular samples with lengths in the range of 0.5 -- 2 mm, widths about 0.5 mm and thicknesses in the range of 0.03 -- 0.1 mm. The potential electrodes where 0.1 x 0.1 mm$^{2}$ pads located from each other at a distance of 0.5, 1.0 or 1.5 mm in dependence on the crystal size The distance between the electrodes is well defined because electrodes are sputtered with a fine mechanical mask. The characteristic ratio of the distance between the Hall contacts to the distance between the current electrodes is about 3 which provides a negligible shortening of the Hall potential  \cite{jan1957galvamomagnetic}.

The dimensions of the crystals were measured using a Zeiss optical microscope and AxioVision software. The uniformity of the thickness was checked by direct observation under the microscope and the thickness values used to calculate the resistivity were obtained from the mass of the crystal and the lateral area of the crystal. For thin crystals this method still gives a large error in absolute values of resistivity and, consequently, in the absolute values of the carrier concentrations determined by the three-band model.

%
\section{Results}
The growth method used provides good homogeneity of the properties of crystals in one batch, while the crystals from different batches can differ significantly, possibly due to deviations of the growth conditions from the optimal ones. Several crystals of FeSe${}_{1-x}$S${}_{x}$ with x=0, 0.037, 0.048, 0.09, and 0.014 have been studied. For unsubstituted FeSe, we studied the crystals of two different batches. The magnetoresistance (MR) at 15~K for these batches is six times different. Apparently, crystals with a lower MR have much more defects \cite{PhysRevB.94.024526}. Their critical temperature is lower than the usual values for FeSe. We studied the properties of this low-quality composition to reveal how the properties of carriers depending on the crystal quality. The batch with x=0 and low MR is further referred to as ``imperfect'' FeSe. On the graphs, the points related to the crystals from this batch are indicated by the orange diamonds.

\begin{figure}[ht]
\includegraphics[scale=0.5,angle=0]{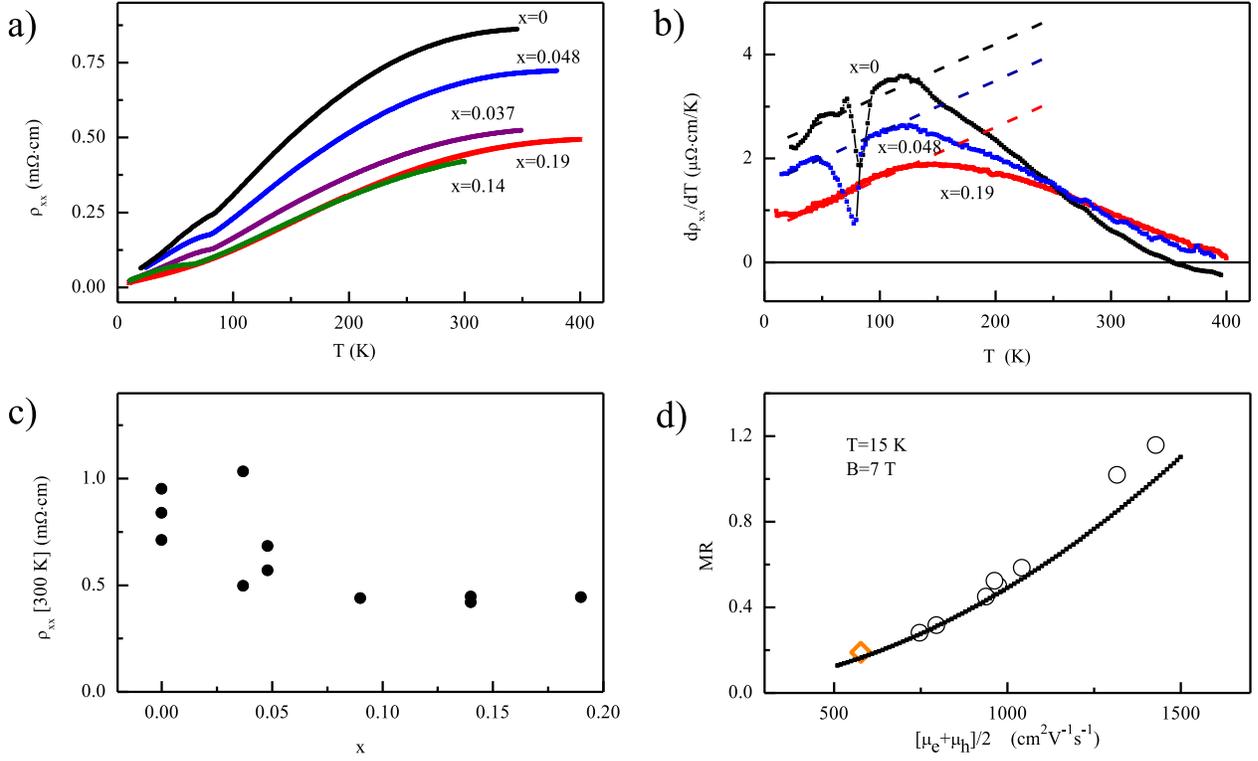}
\caption{ Temperature dependence of in-plane resistivity R (a) and temperature derivative of R [dR/dT] (b) for selected FeSe${}_{1-x}$S${}_{x}$  single crystals. The value of resistivity at room temperature R${}_{300}$ (c) and magnetoresistance at 15~K in 7~T ( MR=[R(B)-R(0)]/R(0) ) (d) for all studied crystals. All the dashed lines in (b) have a slope of  10$^{-8}$  $\Omega$cmK$^{-2}$. The curve in (d) shows $\mu^{2}$B$^{2}$ dependence at B=7~T (the diamond symbol is for ``imperfect'' FeSe). Data for x=0.19 in (a-c) are taken from Ref. \cite{JLTP}.  }
\label{fgr:fig1}
\end{figure}

Figure~\ref{fgr:fig1}(a) shows R(T) curves for the studied samples along with the corresponding curve for FeSe${}_{0.71}$S${}_{0.19}$. The last composition serves as a reference which does not show a transition to an orthorhombic structure at low temperatures \cite{JLTP}. The temperature derivative ( $dR/dT$ ) is plotted in Fig. \ref{fgr:fig1}(b) (for clarity, only three curves ). $dR/dT$ indicates a crossover in transport properties at temperatures in the range of 100-150~K. At low temperatures, $dR/dT$ increases with temperature almost linearly with the exception of the region near the structural transition. This transition causes a singularity on $dR/dT$(T) with the shape of an inverted lambda and, in general, does not change the low-temperature behavior of resistivity. The linear behavior of the derivative of the resistance $dR/dT=aT+b$ yields $aT^{2} +bT+R_{0}$ for R(T) in this temperature range. For all the samples studied, the coefficient $a$ is of the order of 10$^{-8}$  $\Omega$cmK$^{-2}$ which is close to the value reported for LiFeAs  \cite{PhysRevB.84.064512} and, accordingly, should provide a comparable Kadowaki-Woods ratio.

The origin of the crossover in R(T) for the FeSe family is not clear. Saturation can be an important factor because of the high resistance of the compounds at room temperature (0.5 -1.0 m$\Omega$cm) which is close to the Ioffe-Regel limit \cite{PhysRevB.87.024504}. On the other hand, an activation term is clearly detected in the conductivity of FeSe at higher temperatures \cite{karlsson2015study}.

The resistance at room temperature shows a tendency to decrease with increasing $x$ (see Fig.~\ref{fgr:fig1}(c)) although the decrease is not large and the magnitude of the resistivity changes is comparable to the possible errors.

In the magnetotransport properties of all the crystals studied, we observed the features that indicate the presence of carriers with significantly different mobilities. This is a nonlinear field dependence of $\rho_{xy}$ and also a deviation from the simple square law for MR. For a characterization of the carrier transport, we fitted the field dependence of the components of the conductivity tensor, measured at 15~K, with quasiclassical three-band expressions. Details of the method used have been described elsewhere \cite{2017ovchenkovMISM, SUST-30-3-035017}. All the data obtained for the crystals studied are listed in Table~\ref{tbl:T1}.

Figure~\ref{fgr:fig1}(d) shows a relation between magnetoresistance, measured at 15~K in 7~T, and the averaged mobility of the main two bands. The curve in the graph is $\mu^{2}$B$^{2}$ dependence which gives MR of the simplest two-band model with an equal carrier concentration and carrier mobility in the bands. This dependence surprisingly well describes the experimental data, with the exception of two FeSe samples, where we assume a more significant contribution to MR from the mobile carriers. This result agrees with the symmetry of the main electron and hole bands in FeSe, confirmed by high-field measurements \cite{2017ovchenkovMISM}. For a compensated semimetal, such symmetry, most likely, means a triviality of the carriers properties - equal masses and equal scattering times.

\begin{figure}[ht]
\includegraphics[scale=0.5,angle=0]{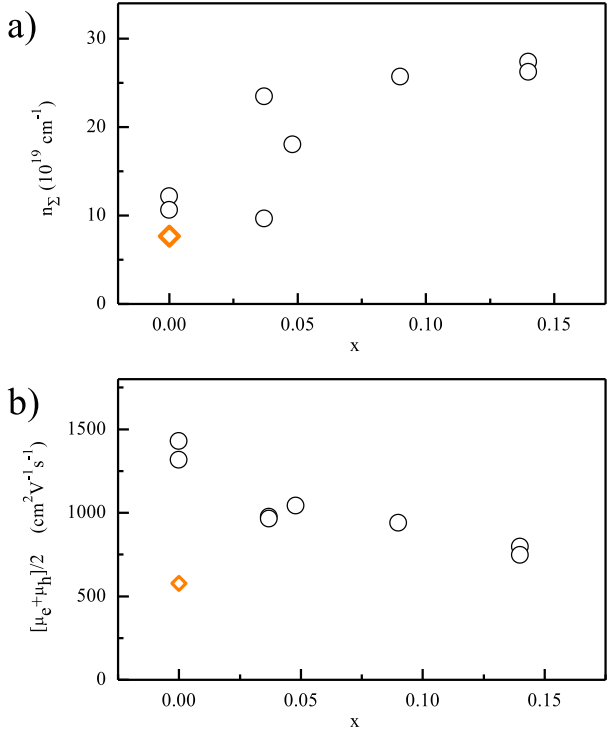}
\caption{ Results of quasiclassical analysis of  $\rho{}_{xx}$(B) and $\rho{}_{xy}$(B) at 15~K and magnetic field up to 7~T using the three-band model for FeSe${}_{1-x}$S${}_{x}$ single crystals versus x. (a) The total carrier concentration n${}_{\sum}$ =[n${}_{e1}$+n${}_{h1}$+n${}_{e2}$]. (b) The average mobility of the main bands [$\mu_{e1}$+$\mu_{h1}$]/2. The diamond symbols are for ``imperfect'' FeSe.  }
\label{fgr:fig2}
\end{figure}

Figure \ref{fgr:fig2} shows the variation of parameters of the main electron and hole bands with $x$. The total carrier concentration and the average mobility are plotted in panels (a) and (b) respectively. The mobility decreases rapidly with increasing degree of substitution, which can be easily explained by the increase in disorder. The carrier mobility in the "imperfect" pure FeSe composition is also very low, which confirms that the change in mobility is mainly caused by the disorder, and not by the changes in microscopic properties. The increase in conductivity during substitution is provided by an increase in carrier concentration shown in the panel (a). Since the substitution is isovalent, the main reason should be a change of the cell parameters which means that the band structure of these compounds is highly sensitive to pressure and stresses. The observed change in electronic properties may be analogous to the transition from bad to good metal in BaFe${}_{2}$(As${}_{1-x}$P${}_{x}$)${}_{2}$ induced by isovalent P substitution \cite{PhysRevB.88.094501}.

\begin{figure}[ht]
\includegraphics[scale=0.5,angle=0]{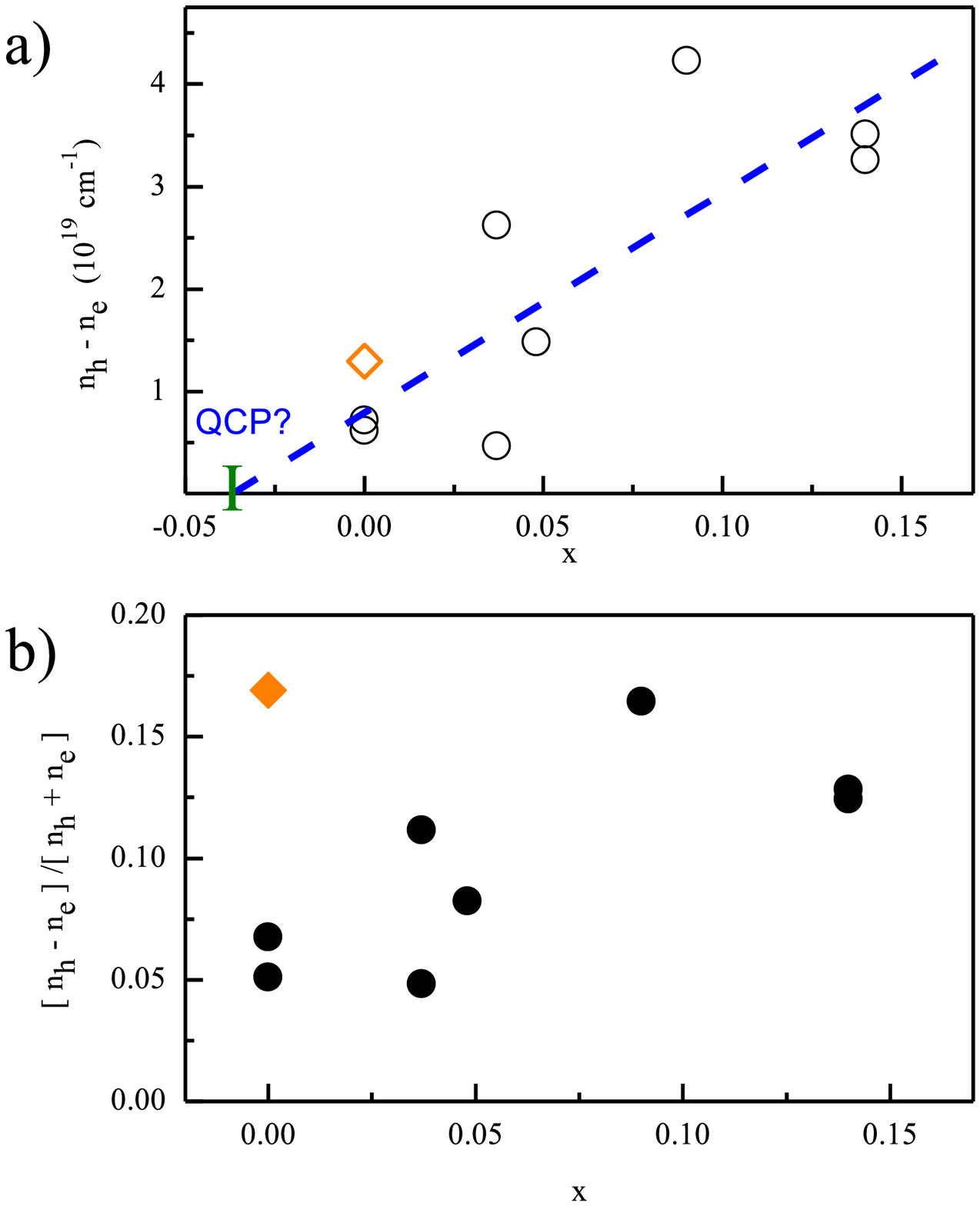}
\caption{ (a) Deviation from full compensation in absolute (a) and relative (b) units. The dashed line is a linear regression. The diamond symbols are for ``imperfect'' FeSe.  }
\label{fgr:fig3}
\end{figure}

Figure \ref{fgr:fig3} shows the variation of the carriers imbalance for the main bands. We do not overestimate the precision of the extracted band parameters and their correspondence to the microscopic properties. Nevertheless, the dependence deserves the closest attention. The observed dependence predicts a full compensation at $x=-0.037$ which may be considered as a tellurium substitution for selenium. We consider the inherent phase separation of FeSe${}_{1-x}$Te${}_{x}$ with a low Te content as a possible experimental evidence of the quantum criticality near the predicted $x$.   Indeed, the structural instability near the carrier-type inversion point can mean that the inversion is not just a numerical coincidence, but a consequence of some qualitative changes in the electronic structure. The proximity of FeSe to quantum criticality has been confirmed by many different experimental methods including the recent muon spin rotation measurements \cite{PhysRevB.97.201102}.

\begin{figure}[ht]
\includegraphics[scale=0.5,angle=0]{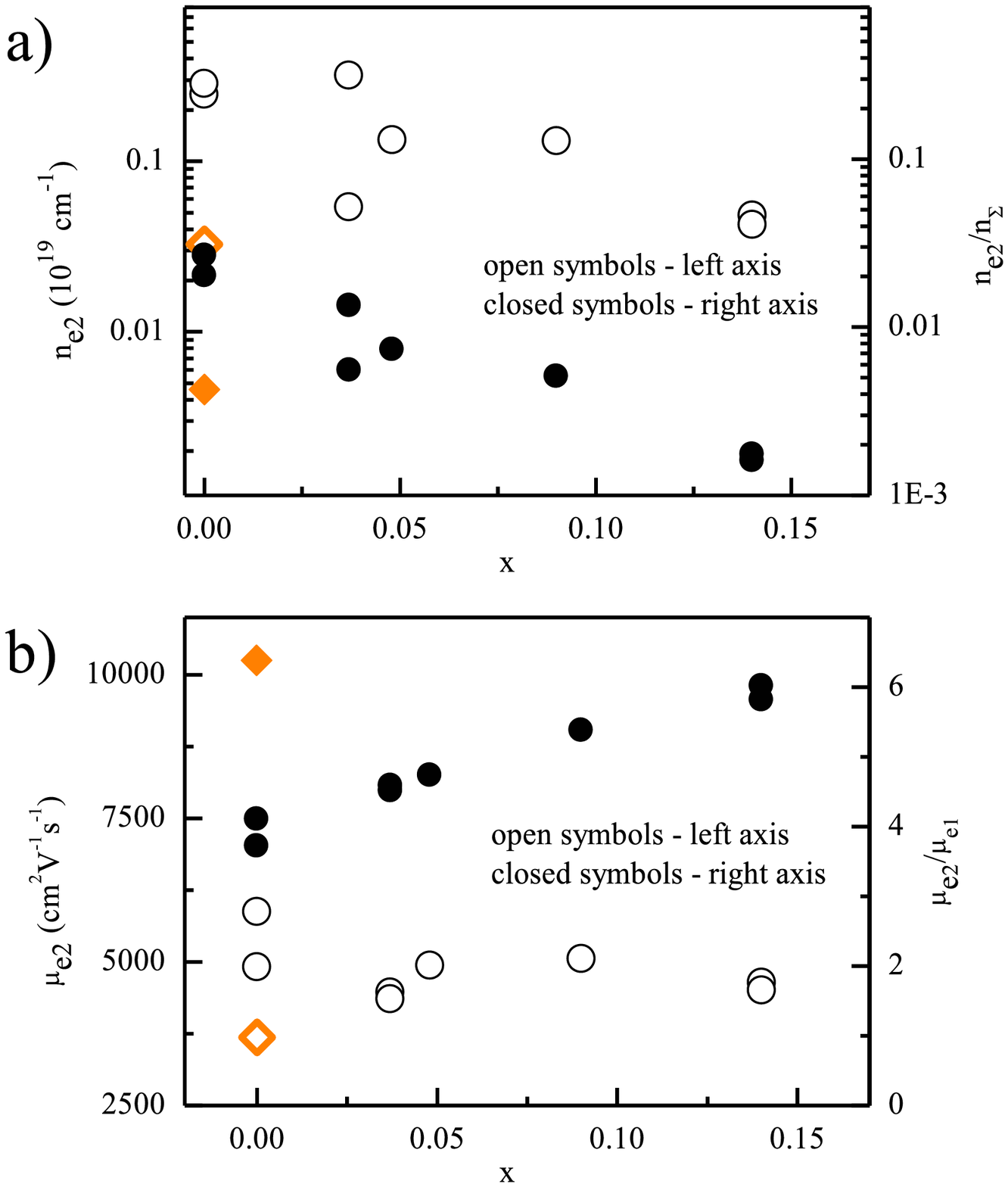}
\caption{ (a) Mobile carriers concentration n$_{e2}$ and the ratio [n$_{e2}$/n${}_{\sum}$]. (b)  Mobile carriers mobility $\mu_{e2}$ and the ratio [$\mu_{e2}$/$\mu_{e1}$]. The diamond symbols are for ``imperfect'' FeSe.  }
\label{fgr:fig4}
\end{figure}

Figures \ref{fgr:fig4}(a) and \ref{fgr:fig4}(b) show the properties of mobile carriers in FeSe${}_{1-x}$S${}_{x}$ series as a function of $x$. The figures show both the absolute values ​​of concentration and mobility, as well as their ratio to the corresponding values of the main carriers. The graphs indicate some important properties of mobile carriers. First, the concentration of mobile carriers in all studied compounds does not exceed 3 percents. This shows the local nature of these carriers. The data in the Table~\ref{tbl:T1} show no correlation between the thickness of the sample and the number of mobile carriers, which means that these carriers are not the surface electronic states. It worth discussing that in many reports the use of similar methods brings significantly higher values for the relative fractions of mobile carriers. In Ref. \cite{PhysRevLett.115.027006} this fraction is equal to 12.5~\%. Even higher values were reported. According to our observation, such results were obtained using the three-band model with the requirement for full electron-hole compensation. Perhaps this assumption is incorrect, and the deviation from charge compensation is an important factor for FeSe family, as we discuss below. We use the three-band model with free parameters. At low temperatures, the field range of 7~T allows us to obtain sufficiently accurate values for FeSe. The extracted mobilities of the main electron and hole bands differ only by a few percents, which is confirmed by the data obtained in pulsed magnetic fields up to 50~T \cite{2017ovchenkovMISM}.

Isovalent sulfur substitution in FeSe${}_{1-x}$S${}_{x}$ causes a rapid decrease in the concentration of mobile carriers (see Fig. \ref{fgr:fig4}(a)). We suppose that this reflects a change of the pocket near the Van Hove singularity at the border of the Brillouin zone. This pocket changes when the degeneracy of the Van Hove point at the border of the Brillouin zone is lifted and the Van Hove singularity approaches the Fermi level. Sulfur substitution suppresses the nematic transition which moves the Van Hove singularity upward that changes the pocket size. Other defects in ``imperfect'' FeSe also can have the similar effect due to the local distortion of the tetragonal symmetry and partial lifting of the degeneracy of the Van Hove singularity.

The variation of the mobility in Fig. \ref{fgr:fig4}(b) can be interpreted in two ways. First, the ratio of mobilities increases with increasing $x$ with the exception of ``imperfect'' FeSe, for which it reaches a maximum. On the other hand, the absolute values of mobility do not show monotone changes, and the increase in the ratio is mainly due to a decrease in the mobility of the main carriers.

The stability of the mobility value in the mobile band of the studied series can indicate that the scattering time for this carriers does not depend on the impurity level which can be expected for carriers of a certain type. In particular, this behavior can indicate a high value of the electron-phonon interaction for the studied mobile carriers from which it can be concluded that it is this group of carriers that provides the record values of acoustic absorption observed in the ultrasonic study of FeSe discussed in the introduction.

\begin{figure}[ht]
\includegraphics[scale=0.5,angle=0]{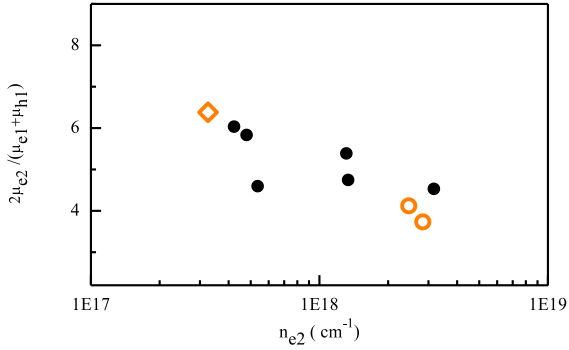}
\caption{ Ratio of mobilities 2$\mu_{e2}$/[$\mu_{e1}$+$\mu_{h1}$] versus  concentration of the mobile carriers. Open symbols are for pure FeSe. }
\label{fgr:fig5}
\end{figure}

Figure\ref{fgr:fig5} shows the ratio of mobilities plotted versus the concentration of the mobile carriers. The linear dependence can be explained by mass renormalization and the change of the pocket size. In our opinion, this confirms that the number of mobile carriers depends on the size of the pocket near the Van Hove singularity and, thus, indicates the position of this point relative to the Fermi level.

\begin{table*}[h]
\caption{\label{tbl:T1}  The list of samples studied with the chemical composition, the thickness of the studied crystal $h$, superconducting critical temperature $T_{C} $, temperature of nematic transition $T_{N} $, MR at 15~K in 7~T, and parameters extracted from magnetotransport measurements at 15~K using three-band model for FeSe${}_{1-x}$S${}_{x}$  single crystals. }
\lineup
\begin{tabular}{cccccccccccc}
\br
& & & & & & \multicolumn{2}{c}{e1}&\multicolumn{2}{c}{h1}&\multicolumn{2}{c}{e2}\\
Sample & $x$ & h & $T_{C} $ & $T_{N} $ & MR & $\mu_{n}$ & $n_{n}$ & $\mu_{p}$ & $n_{p}$ & $\mu_{n}$ & $n_{n}$ \\
& &  \scriptsize{$\mu$ m}&  \scriptsize{K} &   \scriptsize{K} & & \scriptsize{ cm$^{2}$/Vs} &\scriptsize{ 10$^{19}$ cm${}^{-3}$} & \scriptsize{cm$^{2}$/Vs} & \scriptsize{10$^{19}$ cm$^{-3}$} & \scriptsize{cm$^{2}$/Vs} & \scriptsize{10$^{19}$ cm$^{-3}$} \\
 \mr
A0\#1&0&84&9.6&87&1.15 &1470&5.6&1384&6.3  &5871 &0.24\\
A0\#2&0&93&9.5&86&1.02 &1330&4.8&1300&5.5  &4907 &0.28\\
B0\#1&0&99&8.9&84&0.19 &642&3.16&513&4.5  &3687 &0.03\\
A4\#1&0.037&37&10.3&83&0.50&998&4.6&950&5.0  &4473 &0.05\\
A4\#2&0.037&46&10.4&82&0.52&1011&10.2&915 &12.9  &4356  &0.31 \\
A5\#1&0.048&54&10.4&82&0.58&1070&8.2&1014 &9.7  &4939   &0.13 \\
A9\#1&0.09&45&10.3&76&0.44&987&10.7&889 &14.9  &5053   &0.13 \\
A14\#1&0.14&51&10.6&68&0.31&816&11.9&775&15.4 &4636  &0.05\\
A14\#2&0.14&48&10.4&67&0.28&763&11.5&731&14.7  &4502  &0.04 \\
\br
\end{tabular}
\end{table*}

%
\section{Discussion}

Iron-based superconductors are usually considered as semimetals. The type of conductivity in these multiband materials may depend on many factors. Nevertheless, the direct correlation between Hall numbers and critical temperatures was previously observed for  Ba(Fe${}_{1-x}$Co${}_{x}$)${}_{2}$As${}_{2}$ thin films with compressive and tensile in-plane strain in a wide range of Co doping \cite{iida2016hall}. The increase in the critical temperature of FeSe under pressure is also accompanied by changes in Hall numbers and majority carrier-type inversion\cite{PhysRevLett.118.147004}. Our data show that the carrier-type inversion may be an important ingredient of the superconducting phase diagram.

\begin{figure}[ht]
\includegraphics[scale=0.5,angle=0]{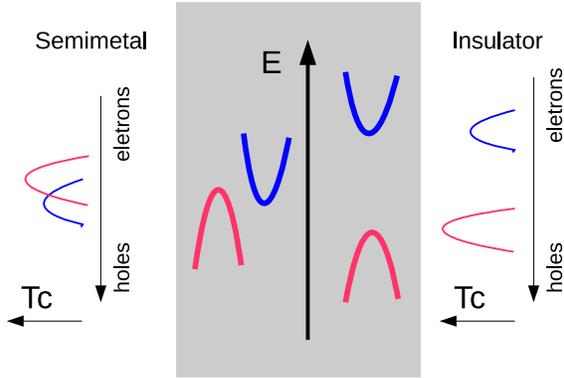}
\caption{Schematic representation of the relationship between the superconducting phase diagram and simplified band structure for iron-based semimetals and cuprate insulators. }
\label{fgr:fig6}
\end{figure}

It is important that a "doped semimetal" model for IBS in conjunction with a doped insulator model for cuprates allow us to give a unified scheme for the formation of a superconducting phase diagram. In Fig.\ref{fgr:fig6} we plot the schematic representation for the relationship between the superconducting phase diagram and band structure in the case of the iron-based semimetals and cuprate insulators. We use the simplest two-band model of the corresponding band structure and suppose that each band provides one superconducting dome in the phase diagram. This phenomenological scheme gives a consistent description of HTSC phase diagram for both IBS and cuprates and explains the origin of the local maxima in phase diagrams of IBS. In the case of FeSe family these local maxima occur for FeSe${}_{1-x}$Te${}_{x}$ at about x=0.5 with maximum T${}_{c}$ near 14~K and for  FeSe${}_{1-x}$S${}_{x}$ at about x=0.1 with T${}_{c}$ near 12~K. The former maximum is most likely in the region of electronic conductivity, and the later in the hole-like region.     

The point of carrier-type inversion is probably a quantum critical point. Within the framework of the band theory, a non-compensation in simple semimetals can be explained by a band nonparabolicity. In this case, the quantum criticality can be related to a Lifshitz transition. In a localized-state approach, the charge compensation is destroyed by a perceptible non-stoichiometry of the studied compositions. In this case, the position of the point corresponding to the "imperfect" FeSe in  \ref{fgr:fig3} (b) may indicate the largest deviation in stoichiometry. Taking into account that for the series studied there was no direct correlation between stoichiometry and compositions, this inversion may be a consequence of the charge inversion in the selenium vacancy or (in other words) the inversion of the charge transfer between the selenium and iron layers.

 The listed assumptions, as well as many other possible explanations, require further study.

\section{Conclusion}
The Van Hove concept is the most natural way to bind the individual energy band and the dome in the superconducting phase diagram. Not necessarily the maximum of the critical temperature should be achieved when the position of the Fermi level and the singularity coincide. For fluctuations of different types, the optimal position may have an energy gap.

\section{Acknowledgments}
We acknowledge support of Russian Foundation for Basic Research through project 17-29-10007 and 16-29-03266. This work has been supported also by the Russian Ministry of Education and Science of the Russian Federation through NUST «MISiS» grant К2-2017-084 and by the Act 211 of the Government of Russia, contracts 02.A03.21.0004, 02.A03.21.0006 and 02.A03.21.0011 and is performed according to the Russian Government Program of Competitive Growth of Kazan Federal University.  


\section*{References}
 \bibliographystyle{unsrt} 
 \bibliography{FeSeS_var.bib}

\begin{thebibliography}{10}

\bibitem{PhysRevLett.19.1039}
J.~Labb\'e, S.~Bari\ifmmode \check{s}\else \v{s}\fi{}i\ifmmode~\acute{c}\else
  \'{c}\fi{}, and J.~Friedel.
\newblock Strong-coupling superconductivity in ${\mathrm{v}}_{3}x$ type of
  compounds.
\newblock {\em Phys. Rev. Lett.}, 19:1039--1041, Oct 1967.

\bibitem{PhysRevLett.56.2732}
J.~E. Hirsch and D.~J. Scalapino.
\newblock Enhanced superconductivity in quasi two-dimensional systems.
\newblock {\em Phys. Rev. Lett.}, 56:2732--2735, Jun 1986.

\bibitem{EPL-3-1225}
J.~Labbé and J.~Bok.
\newblock Superconductivity in alcaline-earth-substituted la 2 cuo 4 : A
  theoretical model.
\newblock {\em EPL (Europhysics Letters)}, 3(11):1225, 1987.

\bibitem{JETP.46.118}
I.~E. Dzyaloshinskii.
\newblock Maximal increase of the superconducting transition temperature due to
  the presence of van't hoff singularities.
\newblock {\em JETP Letters ( Pis'ma Zh. Eksp. Teor. Fiz.)}, 46(3):118, 1987.

\bibitem{PhysRevB.45.5714}
P.~C. Pattnaik, C.~L. Kane, D.~M. Newns, and C.~C. Tsuei.
\newblock Evidence for the van hove scenario in high-temperature
  superconductivity from quasiparticle-lifetime broadening.
\newblock {\em Phys. Rev. B}, 45:5714--5717, Mar 1992.

\bibitem{ABRIKOSOV200097}
A.A. Abrikosov.
\newblock Theory of high-tc superconducting cuprates based on experimental
  evidence.
\newblock {\em Physica C: Superconductivity}, 341-348:97 -- 102, 2000.
\newblock Materials and Mechanisms of Superconductivity High Temperature
  Superconductors VI.

\bibitem{PhysRevLett.89.076401}
V.~Yu. Irkhin, A.~A. Katanin, and M.~I. Katsnelson.
\newblock Robustness of the van hove scenario for high-${T}_{c}$
  superconductors.
\newblock {\em Phys. Rev. Lett.}, 89:076401, Jul 2002.

\bibitem{PhysRevLett.105.067002}
S.~V. Borisenko, V.~B. Zabolotnyy, D.~V. Evtushinsky, T.~K. Kim, I.~V. Morozov,
  A.~N. Yaresko, A.~A. Kordyuk, G.~Behr, A.~Vasiliev, R.~Follath, and
  B.~B\"uchner.
\newblock Superconductivity without nesting in lifeas.
\newblock {\em Phys. Rev. Lett.}, 105:067002, Aug 2010.

\bibitem{2017_Coldea}
Amalia~I. Coldea and Matthew~D. Watson.
\newblock The key ingredients of the electronic structure of fese.
\newblock {\em Annual Review of Condensed Matter Physics}, 9(1):125--146, 2018.

\bibitem{Bohmer2018}
Anna~E B{\"o}hmer and Andreas Kreisel.
\newblock Nematicity, magnetism and superconductivity in fese.
\newblock {\em Journal of Physics: Condensed Matter}, 30(2):023001, 2017.

\bibitem{epl_101-5-56005}
G.~A. Zvyagina, T.~N. Gaydamak, K.~R. Zhekov, I.~V. Bilich, V.~D. Fil, D.~A.
  Chareev, and A.~N. Vasiliev.
\newblock Acoustic characteristics of fese single crystals.
\newblock {\em EPL (Europhysics Letters)}, 101(5):56005, 2013.

\bibitem{PhysRevLett.117.157003}
H.~C. Xu, X.~H. Niu, D.~F. Xu, J.~Jiang, Q.~Yao, Q.~Y. Chen, Q.~Song,
  M.~Abdel-Hafiez, D.~A. Chareev, A.~N. Vasiliev, Q.~S. Wang, H.~L. Wo,
  J.~Zhao, R.~Peng, and D.~L. Feng.
\newblock Highly anisotropic and twofold symmetric superconducting gap in
  nematically ordered ${\mathrm{fese}}_{0.93}{\mathrm{s}}_{0.07}$.
\newblock {\em Phys. Rev. Lett.}, 117:157003, Oct 2016.

\bibitem{PhysRevB.80.140508}
Lei Fang, Huiqian Luo, Peng Cheng, Zhaosheng Wang, Ying Jia, Gang Mu, Bing
  Shen, I.~I. Mazin, Lei Shan, Cong Ren, and Hai-Hu Wen.
\newblock Roles of multiband effects and electron-hole asymmetry in the
  superconductivity and normal-state properties of
  ba(fe${}_{1-x}$co${}_{x}$)${}_{2}$as${}_{2}$.
\newblock {\em Phys. Rev. B}, 80:140508, Oct 2009.

\bibitem{PhysRevB.90.144516}
K.~K. Huynh, Y.~Tanabe, T.~Urata, H.~Oguro, S.~Heguri, K.~Watanabe, and
  K.~Tanigaki.
\newblock Electric transport of a single-crystal iron chalcogenide fese
  superconductor: Evidence of symmetry-breakdown nematicity and additional
  ultrafast dirac cone-like carriers.
\newblock {\em Phys. Rev. B}, 90:144516, Oct 2014.

\bibitem{SUST-30-3-035017}
Y~A Ovchenkov, D~A Chareev, V~A Kulbachinskii, V~G Kytin, D~E Presnov, O~S
  Volkova, and A~N Vasiliev.
\newblock Highly mobile carriers in iron-based superconductors.
\newblock {\em Superconductor Science and Technology}, 30(3):035017, 2017.

\bibitem{Chareev2016}
D.~A. Chareev.
\newblock General principles of the synthesis of chalcogenides and pnictides in
  salt melts using a steady-state temperature gradient.
\newblock {\em Crystallography Reports}, 61(3):506--511, 2016.

\bibitem{CrystEngComm20.2449}
Dmitriy Chareev, Yevgeniy Ovchenkov, Larisa Shvanskaya, Andrey Kovalskii,
  Mahmoud Abdel-Hafiez, Dan~J. Trainer, Eric~M. Lechner, Maria Iavarone, Olga
  Volkova, and Alexander Vasiliev.
\newblock Single crystal growth{,} transport and scanning tunneling microscopy
  and spectroscopy of fese1-xsx.
\newblock {\em CrystEngComm}, 20:2449--2454, 2018.

\bibitem{jan1957galvamomagnetic}
J-P Jan.
\newblock Galvamomagnetic and thermomagnetic effects in metals.
\newblock {\em Solid State Physics}, 5:1--96, 1957.

\bibitem{PhysRevB.94.024526}
A.~E. B\"ohmer, V.~Taufour, W.~E. Straszheim, T.~Wolf, and P.~C. Canfield.
\newblock Variation of transition temperatures and residual resistivity ratio
  in vapor-grown fese.
\newblock {\em Phys. Rev. B}, 94:024526, Jul 2016.

\bibitem{JLTP}
Y.~A. Ovchenkov, D.~A. Chareev, D.~E. Presnov, O.~S. Volkova, and A.~N.
  Vasiliev.
\newblock Superconducting properties of fese\_ $\{$1-x$\}$ s\_ $\{$x$\}$
  crystals for x up to 0.19.
\newblock {\em J. Low Temp. Phys.}, 185:467--473, 2016.

\bibitem{PhysRevB.84.064512}
O.~Heyer, T.~Lorenz, V.~B. Zabolotnyy, D.~V. Evtushinsky, S.~V. Borisenko,
  I.~Morozov, L.~Harnagea, S.~Wurmehl, C.~Hess, and B.~B\"uchner.
\newblock Resistivity and hall effect of lifeas: Evidence for electron-electron
  scattering.
\newblock {\em Phys. Rev. B}, 84:064512, Aug 2011.

\bibitem{PhysRevB.87.024504}
Lev~P. Gor'kov and Gregory~B. Teitel'baum.
\newblock Dual role of $d$ electrons in iron pnictides.
\newblock {\em Phys. Rev. B}, 87:024504, Jan 2013.

\bibitem{karlsson2015study}
S~Karlsson, P~Strobel, A~Sulpice, C~Marcenat, M~Legendre, F~Gay, S~Pairis,
  O~Leynaud, and P~Toulemonde.
\newblock Study of high-quality superconducting fese single crystals: crossover
  in electronic transport from a metallic to an activated regime above 350 k.
\newblock {\em Superconductor Science and Technology}, 28(10):105009, 2015.

\bibitem{2017ovchenkovMISM}
Y.A. Ovchenkov, D.A. Chareev, V.A. Kulbachinskii, V.G. Kytin, D.E. Presnov,
  Y.~Skourski, O.S. Volkova, and A.N. Vasiliev.
\newblock Magnetotransport properties of fese in fields up to 50 t.
\newblock {\em Journal of Magnetism and Magnetic Materials}, 459:221 -- 225,
  2018.

\bibitem{PhysRevB.88.094501}
M~Nakajima, T~Tanaka, S~Ishida, K~Kihou, CH~Lee, A~Iyo, T~Kakeshita, H~Eisaki,
  and S~Uchida.
\newblock Crossover from bad to good metal in bafe 2 (as 1- x p x) 2 induced by
  isovalent p substitution.
\newblock {\em Physical Review B}, 88(9):094501, 2013.

\bibitem{PhysRevB.97.201102}
V.~Grinenko, R.~Sarkar, P.~Materne, S.~Kamusella, A.~Yamamshita, Y.~Takano,
  Y.~Sun, T.~Tamegai, D.~V. Efremov, S.-L. Drechsler, J.-C. Orain, T.~Goko,
  R.~Scheuermann, H.~Luetkens, and H.-H. Klauss.
\newblock Low-temperature breakdown of antiferromagnetic quantum critical
  behavior in fese.
\newblock {\em Phys. Rev. B}, 97:201102, May 2018.

\bibitem{PhysRevLett.115.027006}
M.~D. Watson, T.~Yamashita, S.~Kasahara, W.~Knafo, M.~Nardone, J.~B\'eard,
  F.~Hardy, A.~McCollam, A.~Narayanan, S.~F. Blake, T.~Wolf, A.~A. Haghighirad,
  C.~Meingast, A.~J. Schofield, H.~v.~L\"ohneysen, Y.~Matsuda, A.~I. Coldea,
  and T.~Shibauchi.
\newblock Dichotomy between the hole and electron behavior in multiband
  superconductor fese probed by ultrahigh magnetic fields.
\newblock {\em Phys. Rev. Lett.}, 115:027006, Jul 2015.

\bibitem{iida2016hall}
Kazumasa Iida, Vadim Grinenko, Fritz Kurth, Ataru Ichinose, Ichiro Tsukada,
  Eike Ahrens, Aurimas Pukenas, Paul Chekhonin, Werner Skrotzki, Angelika
  Teresiak, Ruben H\"uhne, Saicharan Aswartham, Sabine Wurmehl, Ingolf M\"onch,
  Manuela Erbe, Jens H\"anisch, Bernhard Holzapfel, Stefan-Ludwig Drechsler,
  and Dmitri~V. Efremov.
\newblock Hall-plot of the phase diagram for ba (fe1- xcox) 2as2.
\newblock {\em Scientific Reports}, 6:28390, 2016.

\bibitem{PhysRevLett.118.147004}
J.~P. Sun, G.~Z. Ye, P.~Shahi, J.-Q. Yan, K.~Matsuura, H.~Kontani, G.~M. Zhang,
  Q.~Zhou, B.~C. Sales, T.~Shibauchi, Y.~Uwatoko, D.~J. Singh, and J.-G. Cheng.
\newblock High-${T}_{c}$ superconductivity in fese at high pressure: Dominant
  hole carriers and enhanced spin fluctuations.
\newblock {\em Phys. Rev. Lett.}, 118:147004, Apr 2017.

\end{thebibliography}


\section{SUPPLEMENTARY MATERIAL}

\subsection{Method}

For the multicarrier analysis we used a quasiclassical relaxation-time approximation for the field dependence of the conductivity tensor components of a layered tetragonal crystal as a sum of $l$ band terms \cite{jan1957galvamomagnetic}. The analysis is based on fitting the experimental data with three-band model. First, we calculate the conductivity components from the measured resistivity components. For tetragonal crystals:
\begin{eqnarray}
\sigma_{xx}=\sigma_{yy}=\frac{\rho_{xx}}{(\rho_{xx}^{2}+\rho_{xy}^{2})} \nonumber\\
-\sigma_{xy}=\sigma_{yx}=\frac{\rho_{xy}}{(\rho_{xx}^{2}+\rho_{xy}^{2})}
\end{eqnarray}

where $\sigma_{ij}$ are the conductivity tensor components and $\rho_{ij}$ are the resistivity tensor components

The conductivities of bands are additive, and within the relaxation-time approximation for an arbitrary number of bands, we can write
\begin{eqnarray}
\sigma_{xx}=F_{R}(B)\equiv\sum_{i=1}^{l}\frac{|\sigma_{i}|}{(1+\mu_{i}^{2}B^2{})}\nonumber\\
\sigma_{xy}=F_{H}(B)\equiv\sum_{i=1}^{l}\frac{\sigma_{i}\mu_{i}B}{(1+\mu_{i}^{2}B^2{})}\nonumber\\
\sigma_{i}=en_{i}\mu_{i}
\end{eqnarray}

where $i$ is the band index, $e$ is the electron charge, $\sigma_{i}$ is the conductivity at $B=0$, $\mu_{i}$ is the mobility, $n_{i}$ is the carrier concentration, and $l$ is the number of bands. The fitting procedure determines $\mu_{i}$ and $n_{i}$ by minimizing the residual $\phi$:
\begin{equation}
\phi=\frac{1}{N}\sum_{k=1}^{N}{\bigg[\bigg({\frac{\sigma_{xx}[k]-  F_{R}(B[k])}{\sigma_{xx}[k]}\bigg)}^{2}+\bigg({\frac{\sigma_{xy}[k]- F_{H}(B[k])}{\sigma_{xy}[k]}\bigg)}^{2} \bigg]} \nonumber
\end{equation}
where $\sigma_{xx}[k]$, $\sigma_{xy}[k]$, and $B[k]$ are the values of $\sigma_{xx}$, $\sigma_{xy}$, and $B$ at experimental point $k$, and $N$ is the number of the measured points.

\subsection{Results}

\begin{figure}[ht]
\includegraphics[scale=0.5,angle=0]{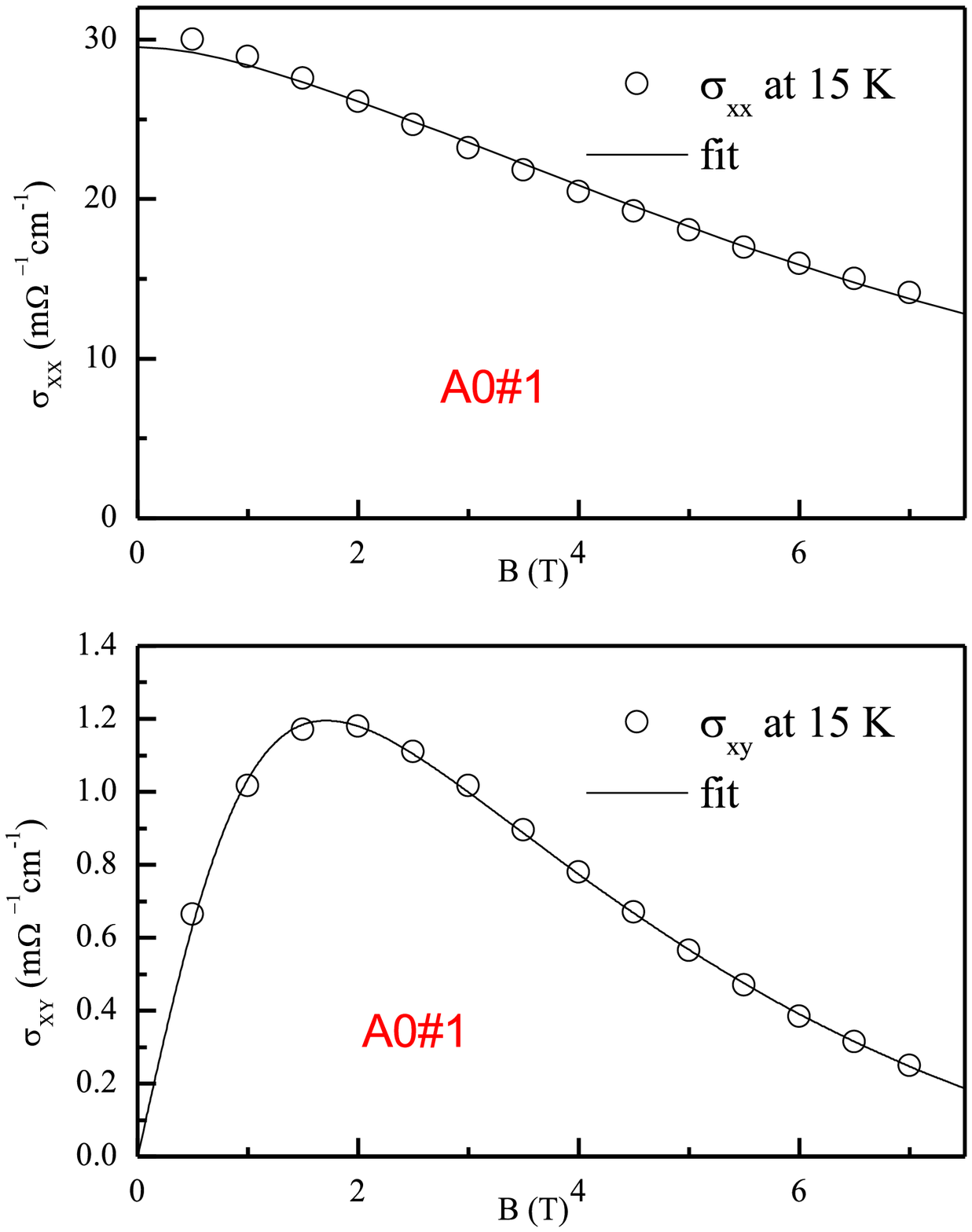}
\caption{ Experimental data and the best three-band fit ($\phi$=2.3$\times$10$^{-4}$). }
\label{fgr:fig_s1}
\end{figure}

\begin{figure}[ht]
\includegraphics[scale=0.5,angle=0]{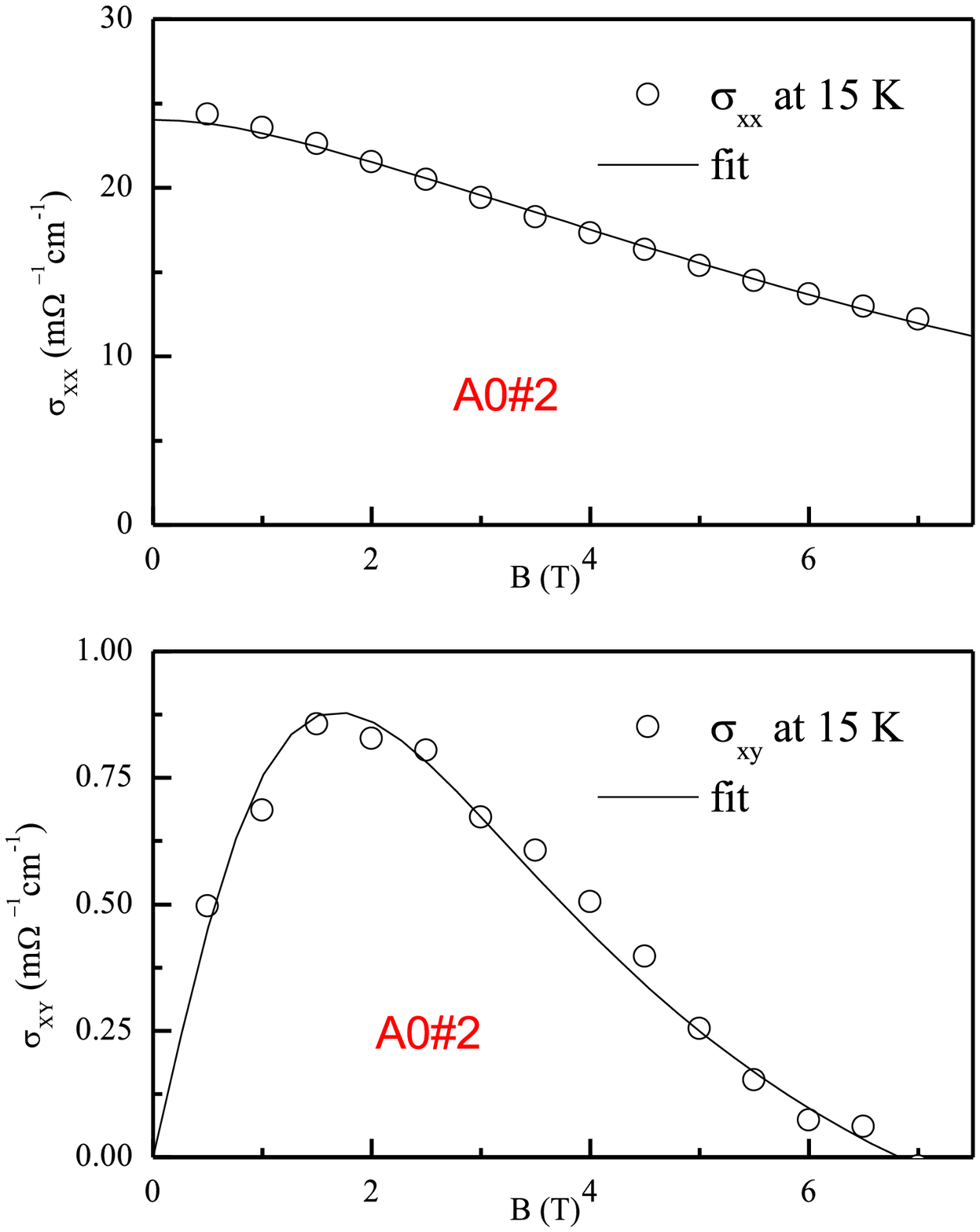}
\caption{ Experimental data and the best three-band  fit ($\phi$=1.3$\times$10$^{-2}$). }
\label{fgr:fig_s2}
\end{figure}

\begin{figure}[ht]
\includegraphics[scale=0.5,angle=0]{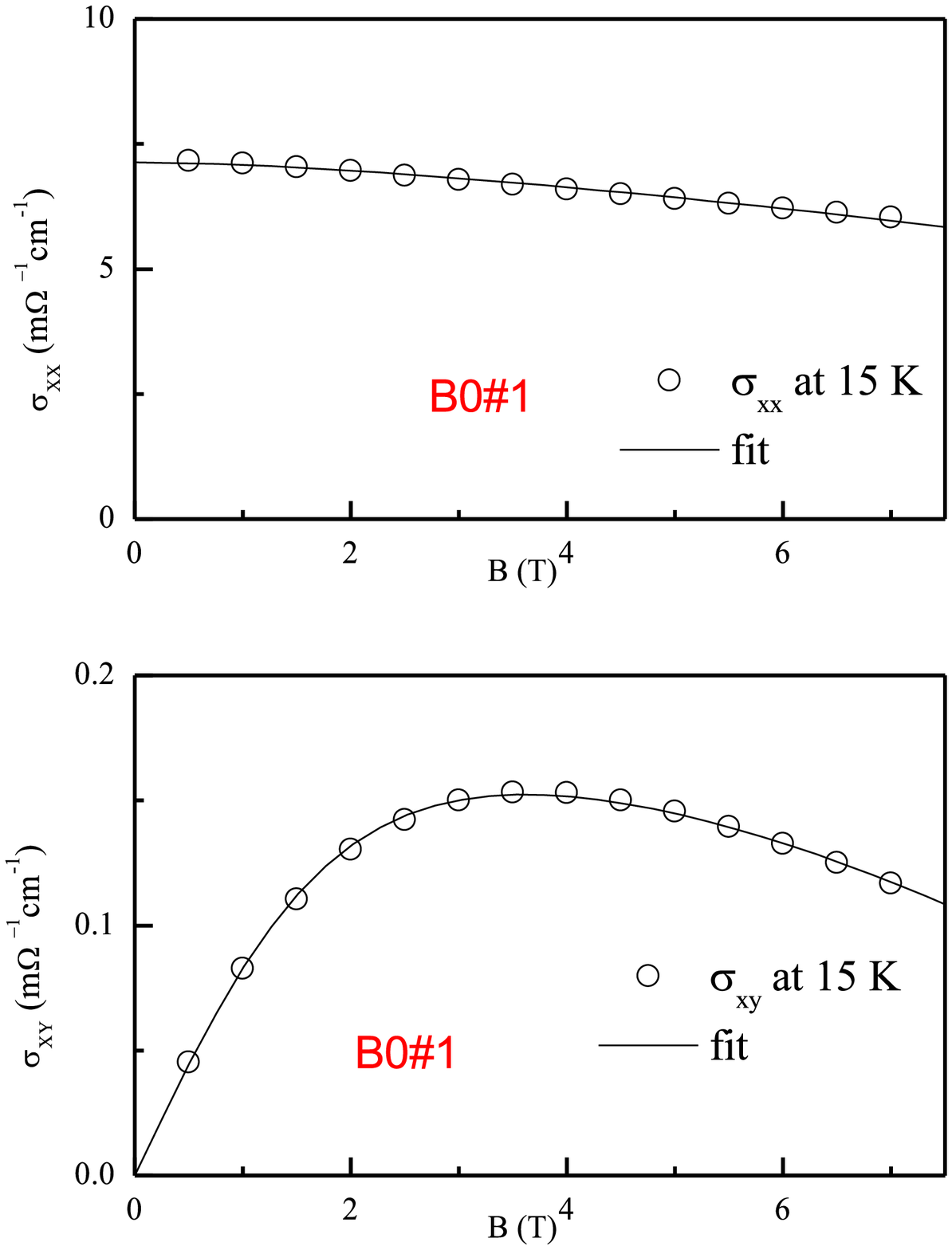}
\caption{ Experimental data and the best three-band  fit ($\phi$=3.5$\times$10$^{-5}$). }
\label{fgr:fig_s3}
\end{figure}

\begin{figure}[ht]
\includegraphics[scale=0.5,angle=0]{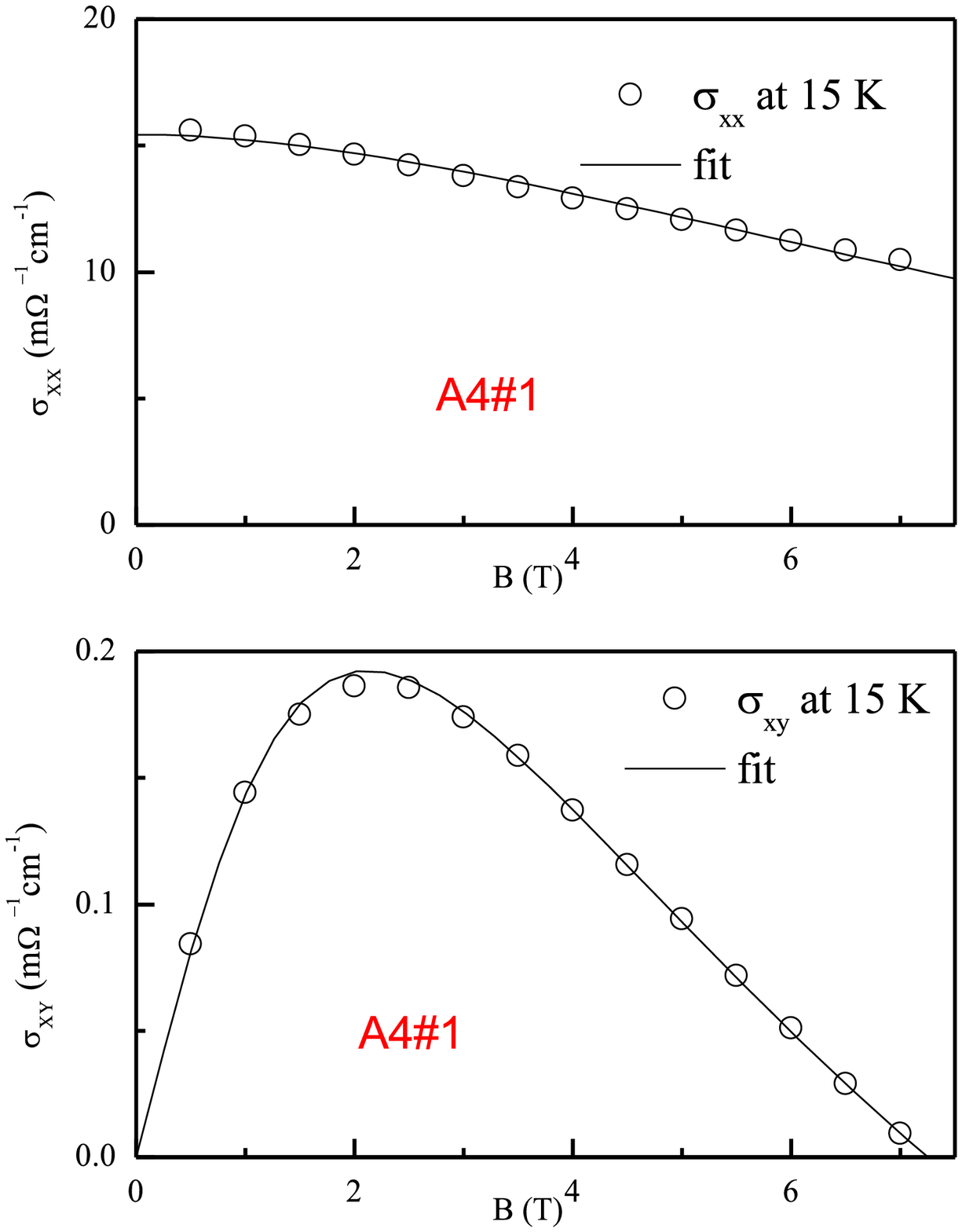}
\caption{ Experimental data and the best three-band  fit ($\phi$=1.3$\times$10$^{-4}$). }
\label{fgr:fig_s4}
\end{figure}

\begin{figure}[ht]
\includegraphics[scale=0.5,angle=0]{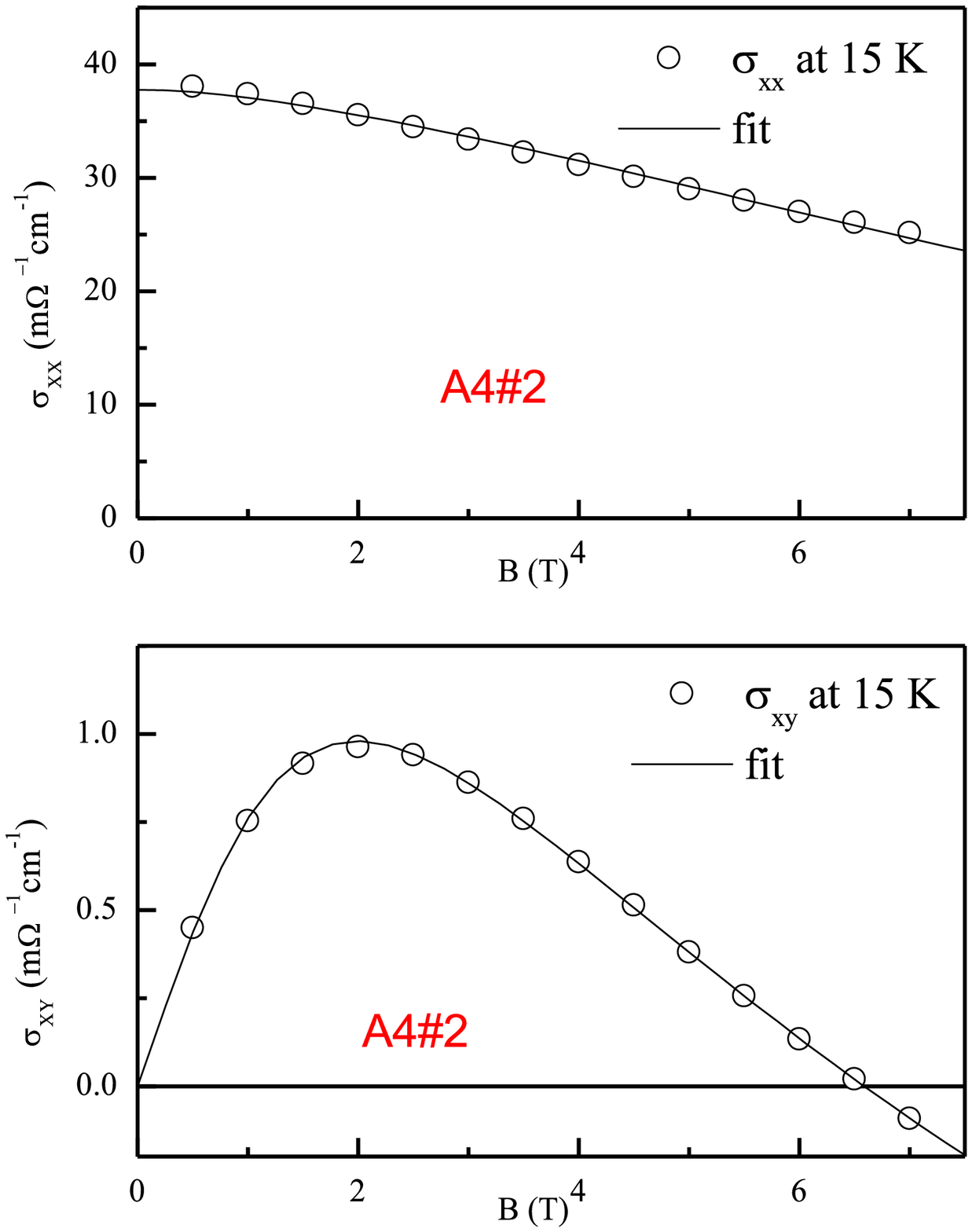}
\caption{ Experimental data and the best three-band  fit ($\phi$=1.2$\times$10$^{-4}$). }
\label{fgr:fig_s5}
\end{figure}

\begin{figure}[ht]
\includegraphics[scale=0.5,angle=0]{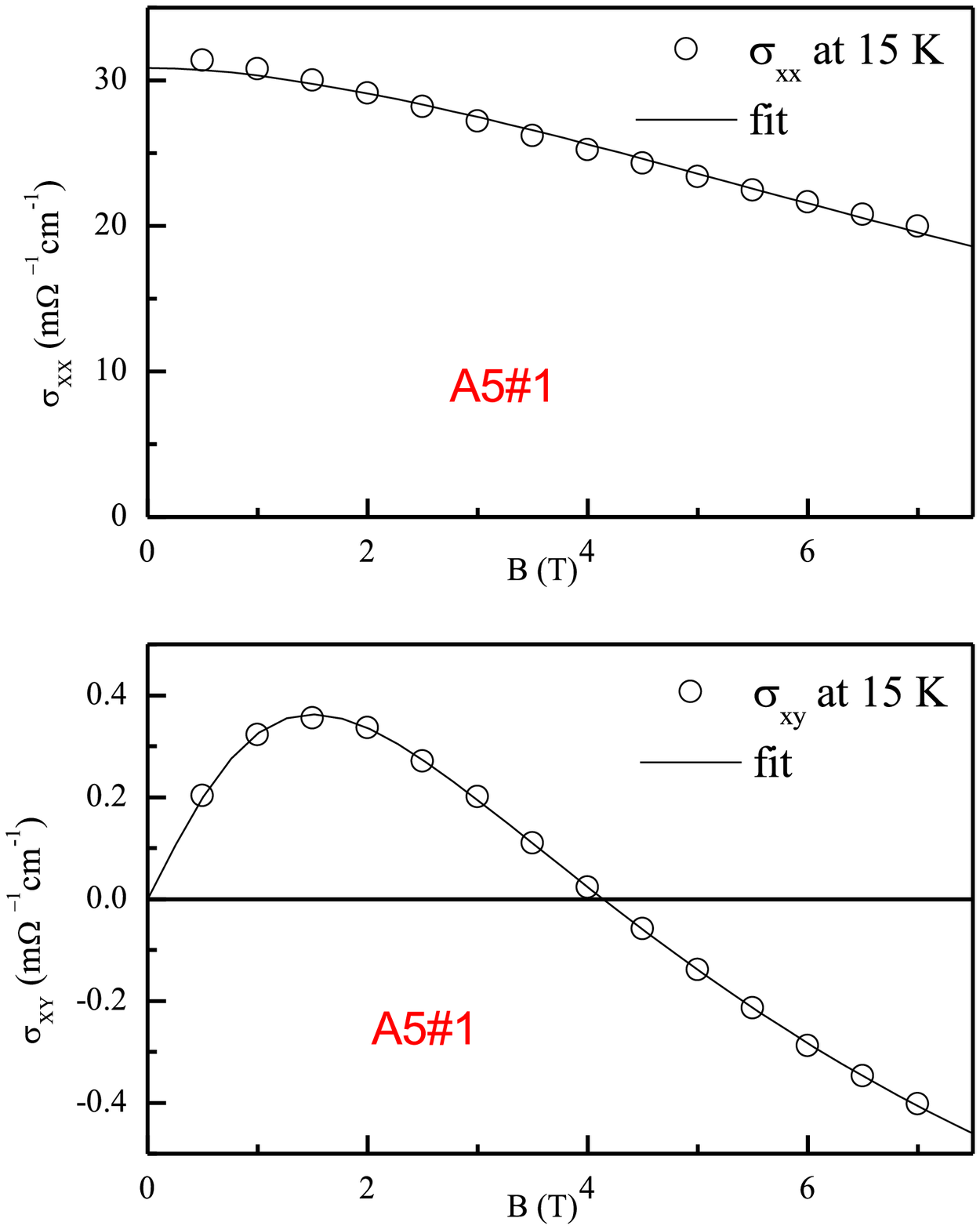}
\caption{ Experimental data and the best three-band  fit ($\phi$=1.7$\times$10$^{-4}$). }
\label{fgr:fig_s6}
\end{figure}

\begin{figure}[ht]
\includegraphics[scale=0.5,angle=0]{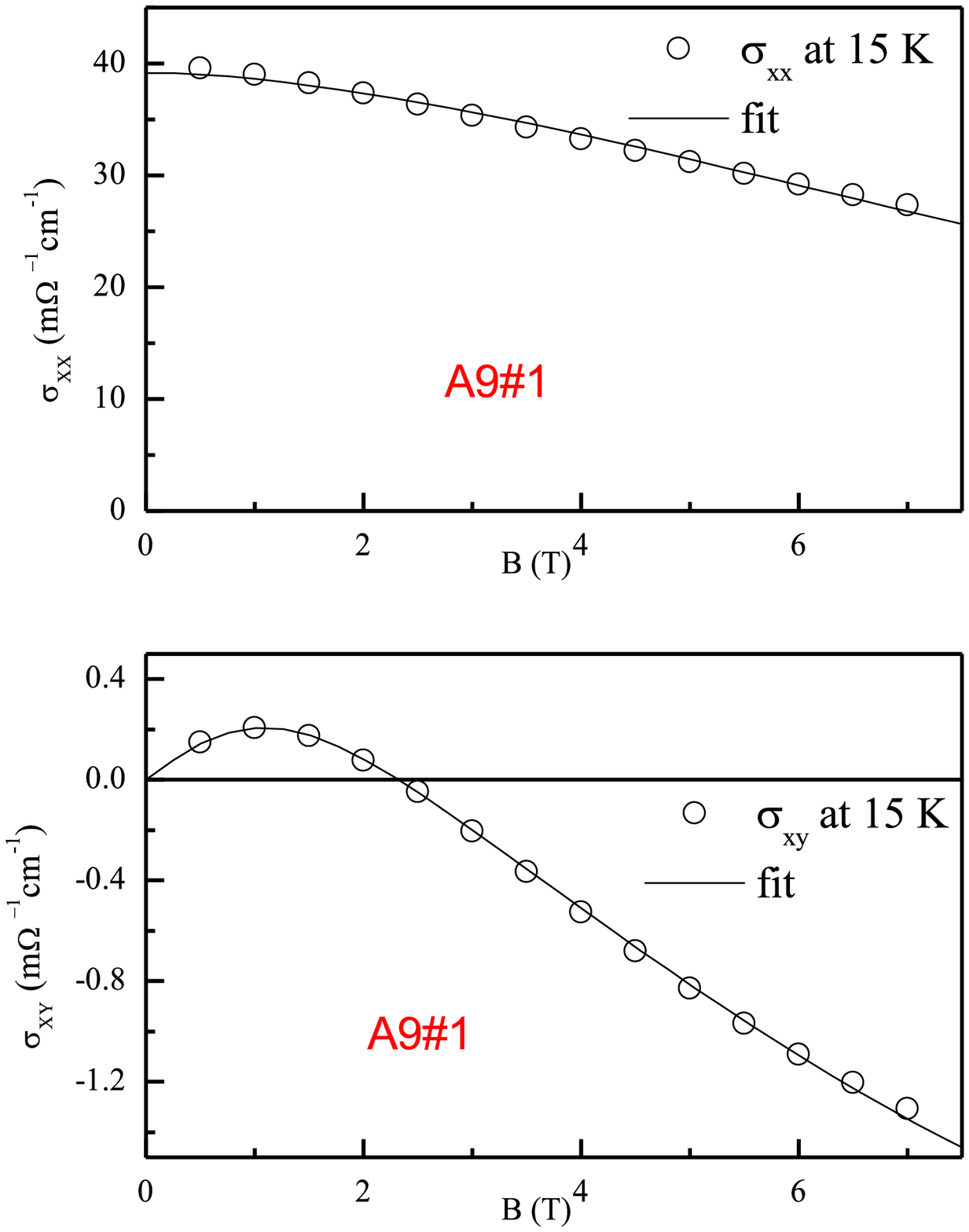}
\caption{ Experimental data and the best three-band  fit ($\phi$=4.6$\times$10$^{-4}$). }
\label{fgr:fig_s7}
\end{figure}

\begin{figure}[ht]
\includegraphics[scale=0.5,angle=0]{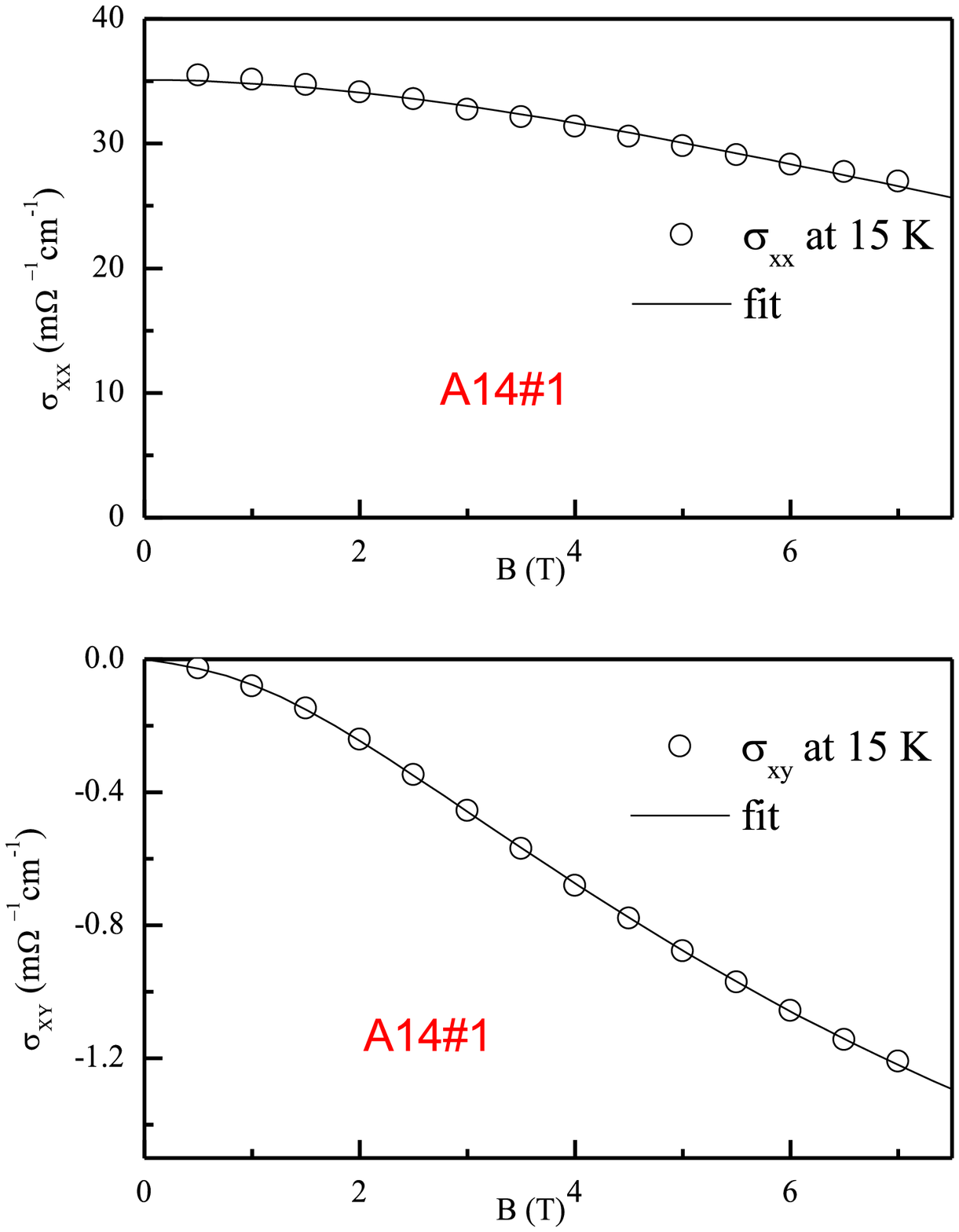}
\caption{ Experimental data and the best three-band  fit ($\phi$=1.8$\times$10$^{-4}$). }
\label{fgr:fig_s8}
\end{figure}

\begin{figure}[ht]
\includegraphics[scale=0.5,angle=0]{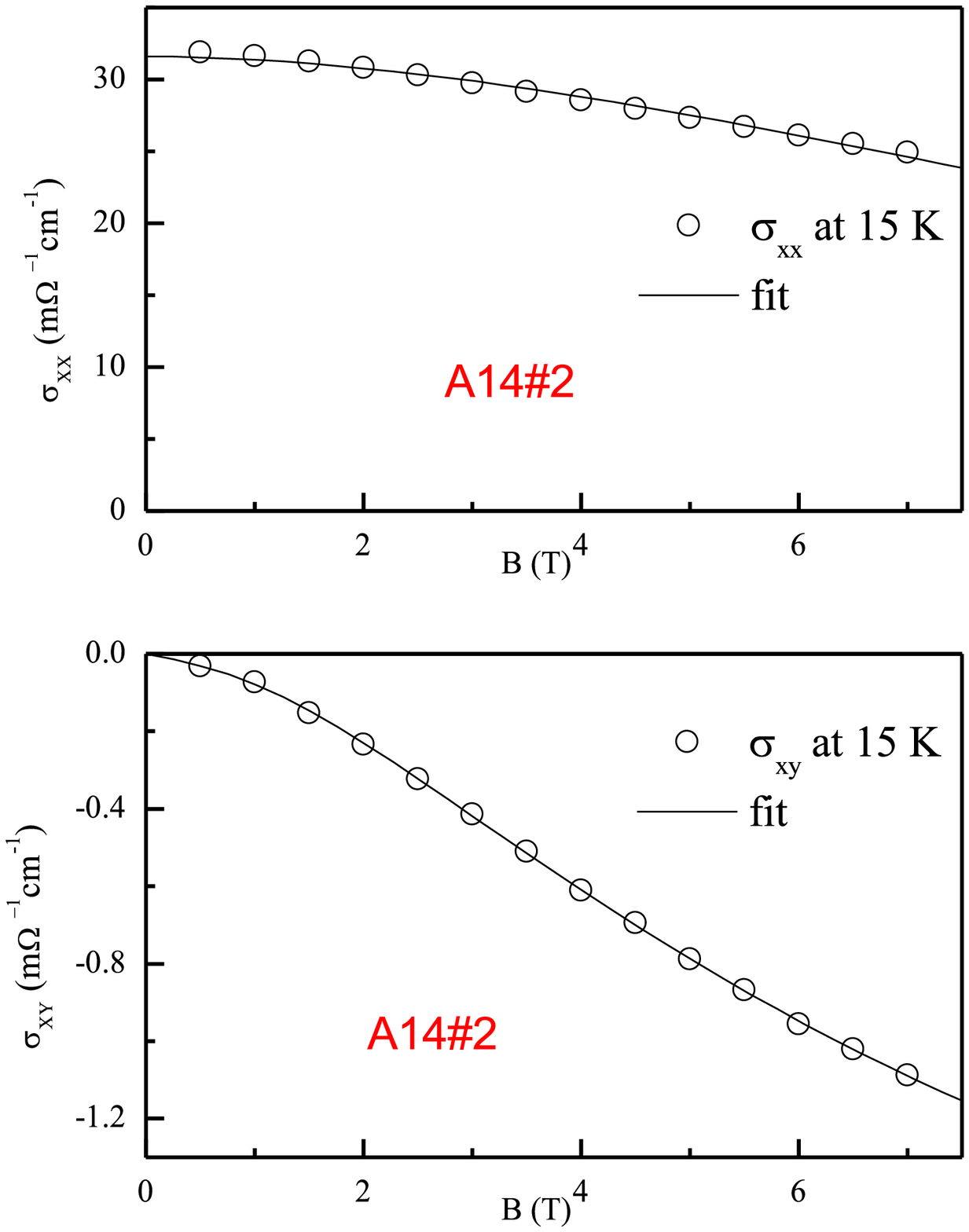}
\caption{ Experimental data and the best three-band  fit ($\phi$=3.1$\times$10$^{-4}$). }
\label{fgr:fig_s9}
\end{figure}


\end{document}